# Evolution of Computer Virus Concealment and Anti-Virus Techniques: A Short Survey

Babak Bashari Rad[1], Maslin Masrom[2] and Suhaimi Ibrahim[3]

[1] Faculty of Computer Science and Information System, University Technology Malaysia
Skudai, 81310 Johor, Malaysia
babak.basharirad@hotmail.com

[2] Razak School of Engineering and Advanced Technology, University Technology Malaysia
Kuala Lumpur, 54100 Selangor, Malaysia
maslin@ic.utm.my

[3] Advanced Informatics School, University Technology Malaysia
Kuala Lumpur, 54100 Selangor, Malaysia
suhaimiibrahim@utm.my

**Abstract**
This paper presents a general overview on evolution of concealment methods in computer viruses and defensive techniques employed by anti-virus products. In order to stay far from the anti-virus scanners, computer viruses gradually improve their codes to make them invisible. On the other hand, anti-virus technologies continually follow the virus tricks and methodologies to overcome their threats. In this process, anti-virus experts design and develop new methodologies to make them stronger, more and more, every day. The purpose of this paper is to review these methodologies and outline their strengths and weaknesses to encourage those are interested in more investigation on these areas.

## 1. Introduction

Since the first days of appearance of early malwares, there is a big contest between virus creators and anti-virus experts and it is becoming more complicated every day, and will continue afterward. At the same time as anti-virus softwares are advancing their methods, on the other hand, the virus writers are seeking for new tactics to overcome them. They utilize various techniques to put their products out of sight of the scanners, because if antivirus programs can easily find their viruses, they cannot sufficiently spread far in the wild. Hence, virus authors always struggle to create new code evolution tactics to beat against the detectors. Accordingly, virus techniques grew increasingly throughout all the years, from plainest methods to some more advanced strategies.

Although, the anti-virus specialists generally follow the new techniques used in advanced malwares and attempt to overcome them, however, all the new defense techniques are not sufficient and there is an extremely necessity for more researches. The most important thing is to analyze and understand all the previous methods, well.

In this paper, firstly, we give a short description on evolution of computer viruses and their classifications in aspect of concealment tactics. Then we survey the most common scanning and detection methods used in anti-virus software. Anti-virus software employed different methodologies in analyzing, scanning, and detecting viruses to provide sufficient safety for computer systems. In the next section, we present a comparison table that shows these detection methods and their features. It helps us to understand the advantages and disadvantages of each method and compare them. In the last section, we terminate with a conclusion and some recommendations.

## 2. Evolutionary Concealment

Computer malwares can be classified according to their different characteristics in several various manners, such as classification by target or classification by infection mechanism. One of these classification types is according to concealment techniques employed.

### 2.1 Encrypted Viruses

Encryption is practically the most primitive approach to take cover the operation of the virus code [1]. The ultimate aim of encrypted viruses is change of the virus body binary codes with some encryption algorithms to hide it from simple view and make it more difficult to analyze and detect [2]. The first encrypting virus, CASCADE, appeared in 1988 [3].

Normally, encrypted viruses are made of two key parts: the encrypted body of the virus, and a small decryption code piece [4]. When the infected program code gets to run, firstly, the decryption loop executes and decrypts the main body of the virus. Then, it moves the control to the virus body. In some viruses, decryption loop performs something more, in addition to its main task. For instance, it may calculate the checksum to make sure that the virus code is not tampered, but as a general principle, the decryptor should be created as small as possible to avoid the anti-virus software, which is trying to exploit the decryptor loop's string pattern for scanning purpose.

Encryption hides the virus body from those who like to view the virus code or tamper the infected files using code viewers or hexadecimal editors [5]. However, virus programmers use the encryption for some reasons. Four of the major motivations as described by Skulason [4] are:

**1. To avoid static code analysis:** Some programs try to analyze code automatically and generate warning if suspect code is found. Encryption is used to disguise suspicious codes and prevent static analysis.

**2. To delay the process of inspection:** It can make the analysis process a bit more difficult and time-consuming, however it usually can increase the time of process only a few minutes.

**3. To prevent tampering:** Many new variants of a virus can be produced with a minor change in the original virus code. Encryption makes it difficult to change the virus by non-experts.

**4. To escape from detection:** an encrypted virus cannot be detected through simple string matching before decryption, because only decryptor loop has identical string in all variants. Hence, signature for an encrypted virus is limited and must be selected precisely.

## 2.2 Oligomorphic Virus

Although virus creators attempted to conceal the first generation of viruses with encryption methods, the decryptor loops were remained constantly in new infected files, so anti-virus software normally had no trouble with such virus that was inspected and for which a signature string was obtained. To overcome this vulnerability, virus writers employed several techniques to create a mutated body for decryptors. These efforts caused the invention of new type of concealment viruses, named as oligomorphic viruses.

Oligomorphic viruses are willing to substitute the decryptor code in new offspring. The easiest method to apply this idea is to provide a set of different decryptor loops rather than one. Signature based detection depending on byte pattern of decryptor, though it is a achievable solution, but it is not a practical way [1]. Therefore, oligomorphic viruses make the detection process more difficult for signature based scanning engines.

## 2.3 Polymorphic Virus

The most usual approach developed in anti-virus softwares and tools to identify the viruses and malwares is signature-based scanning [6]. It makes use of small strings, named as signatures, results of manual analysis of viral codes. A signature must only be a sign of a specific virus and not the other viruses and normal programs. Accordingly, a virus would be discovered, if the virus related signatures were found. To avoid this detection, virus can change some instructions in new generation and cheat the signature scanning. Polymorphic viruses exploit this concept. When the virus decides to infect a new victim, it modifies some pieces of its body to look dissimilar. As encryption and oligomorphism, scheme of polymorphism is to divide the code into two sections, the first part is a code decryptor, which its function is decryption of the second part and passes the execution control to decrypted code. Then, during the execution of this second part, a new different decryptor will be created, which encrypts itself and links both divisions to construct a new copy of the virus [7, 8].

In fact, polymorphism is a newer and progressive variety of oligomorphism. Concerning of encryption, polymorphic virus, oligomorphic and encrypted viruses are similar, but the exception is the polymorphic virus has capability to create infinite new decryptors [2, 9]. Polymorphic virus exploits mutation techniques to change the decryptor code. Furthermore, each new decryptor may use several encryption techniques to encrypt the constant virus body, as well.

## 2.4 Metamorphic Virus

Virus writers like to make the lifetime of their produced viruses longer, so they constantly challenge to make the detection as more difficult as possible for antivirus specialists. They have to spend a plenty of time to produce a new polymorphic virus that it may not be able to spread out broadly, but an anti-virus expert may handle the detection of such a virus in a short time [10].

Even for the most complicated polymorphic viruses, after code be emulated sufficiently, the original code will become visible and can be detected by a simple string signature scanning [11].

Peter Szor quoted in [9], the shortest definition of the metamorphic virus, which defined by Igor Muttik, is "Metamorphics are body-polymorphics." Because metamorphic viruses are not encrypted, they do not require decryptor. Metamorphic virus is similar to polymorphic virus in aspect of making use of an obfuscation engine. Metamorphic virus mutate all of its body, rather it changes the code of decryption loop. All possible techniques applicable by polymorphic virus to produce new decryptor can be used by a metamorphic virus on whole virus code to create a new instance.

## 3. Anti-Virus Techniques

### 3.1 First-Generation Scanners

Scanners of first-generation employed not complicated techniques in order to find known computer viruses. Earliest scanners typically looked for certain patterns or sequences of bytes called string signatures.

Once a virus is detected, it can be analyzed precisely and a unique sequence of bytes extracted from the virus code. This string often called signature of the virus and is stored in the anti-virus scanner database. It must be selected such that not likely is appeared in benign programs or other viruses, optimistically. This technique uses this signature to detect the previously analyzed virus. It searches the files to find signatures of the viruses. It is one of the most basic and simplest methods employed by antivirus scanners. Some examples of virus signature strings, which are published in Virus Bulletin [12], are given in Table 1.

Table 1: Examples of viruses string signature

| *Virus Name* | *String Pattern (Signature)* |
|---|---|
| Accom.1280 | 89C3 B440 8A2E 2004 8A0E 2104 BA00 05CD 21E8 D500 BF50 04CD |
| Die.448 | B440 B9E8 0133 D2CD 2172 1126 8955 15B4 40B9 0500 BA5A 01CD |
| Xany.979 | 8B96 0906 B000 E85C FF8B D5B9 D303 E864 FFC6 8602 0401 F8C3 |

The anti-virus engine scans the binary code of files to find these strings; if it encounters with a known pattern, it alerts detection of the matching virus. Normally, a string contain of sixteen unique bytes is properly adequate in size to distinguish a 16-bit viral code with no false positive. However, for 32-bit malicious codes to be recognized accurately, more enlarged strings are required, specifically if the malware is created through high level programming languages [1].

#### 3.1.1 Special Cases in String Scanning

Sometimes signature string scanning need some special conditions in bytes comparison process. Some of most useful exceptional cases in string scanning are [1]:

**(i) Wildcards:** Use of wildcards allows excluding some constant byte values or ranges of values from comparison. For example, in the following string, bytes identified by the '?' symbol are not considered in matching. The wildcard %2 indicates the scanner attempt to find the following byte value in the next two places.

*8B96 09?? B000 E85C %2 FF8B D5B9 D303 E864*

Some of earlier encrypted and polymorphic viruses are detectable by wildcards. Furthermore, W32/Regswap metamorphic virus could be detected using this method [9].

**(ii) Mismatches:** It allows negligible values for non-specified quantity of bytes inside a string, regardless of their position. For example, the string *0A 32 B3 17 80 F6 24* with mismatch *2*, is compatible with these strings:

*13 0A 32 01 17 80 4D 24 72*
*0A D9 1E 17 80 F6 24 AA 92*

**(iii) Generic Degree:** when a virus has more than one variant, the variants are analyzed to extract one unique string that indicates all of them. This kind of string scanning uses this common pattern to find any previously identified variants of a virus family. It often exploits wildcards and mismatches, as well, to cover several different patterns of virus family variants.

#### 3.1.2 Bookmarks

Use of Bookmarks is a simple way to ensure a more reliable detection and decrease the risk of false positive. For example, the number of bytes located between the beginning of the virus code and the first byte of the signature can be use as a suitable bookmark. Choosing a reliable bookmark is virus host specific. For example, for boot viruses, appropriate bookmarks may show addresses of boot sectors placed on the disk. In the file-infecting viruses, a proper bookmark can point to the offset of original program header. In addition, the length of the virus could be a very helpful bookmark.

### 3.1.3 Speed up Techniques

Almost all anti-virus scanners overspend most of the search time matching input data with previously discovered virus signatures. Normally, scanners employ various types of multi signature string comparing algorithm. In 2005, more than 100,000 virus signatures was known and is increasing continuously [13]. Therefore, the algorithms need to be performed as faster as possible. There are several techniques to make the string scanning faster. Some of most common methods to speed up the searching algorithm are:

**Hashing:** Generally, hashing techniques are commonly employed in searching algorithms as data structure to make the access to elements based on nonnumeric or large value keys faster and improve the speed of the process, generally. In anti-virus scanners, hashing is exploited in order to decrease the number of searching strings within the file. The methods may use 1 byte or 16-bit or 32-bit words of the scan strings to produce a hash value, which is the index in the hash table [14].

**Top-and-Tail Scanning:** Because virus codes are sited usually at the beginning or end of the victim files, scanning only the first and the last parts, instead of whole file is a useful idea to raise the speed of signature detection procedure more. This procedure is known as top-and-tail scanning. It reduces the number of disk accesses and optimizes scanning speed. However, since the scanners will search only in some specific areas, it may produce false negative results and diminish the accuracy and reliability [14].

**Entry-Point and Fixed-Point scanning:** These techniques also help the scanning engines to execute more rapidly. They use the concept of the program execution entry-point, which is achievable by the headers of executable files. Because viruses the usually seek entry-point of the file as a target, search can be started from this point. In order to keep the execution of a file in normal manner, the virus has to get the executing control from the original start point and pass it again to the infected file original entry point after it terminates its code subroutine.

Fixed-point scanning is useful when there is no sufficient helpful string in the entry point. The scanner firstly specifies an initial location M. Then it tries to find every string that is compatible with signature, at positions M + X. These scanners reduce considerably the number of Input/Output access on disk and speed up the algorithm running process [1].

### 3.2 Second-Generation Scanners

The second-generation scanners started to develop, when the simple pattern scanning techniques lost their efficiency for detecting newer and more complicated viruses, appropriately. In addition, this generation of scanners introduced exact and almost exact recognition that caused the antivirus scanners became more trustable.

### 3.2.1 Smart Scanning

Smart scanning refers to a defense optimizing method for the newer generation of viruses, which try to conceal their code within a sequence of worthless instructions such as no operation NOP instructions.

When virus-mutating kits started to develop, simple signature-based scanning was not so effectual way because these pre-supplied kits could produce viruses much different visually from their original appearance. The mutation prepared tools can insert junk instructions, which have no effect on the execution process, among the program source instructions.

Smart scanning skip junk instructions, like NOPs, and do not consider them as the virus signature bytes. In addition, to improve the detection possibility of variants of a virus, a region of the virus body is chosen which does not include any addresses of data or other subroutines. Furthermore, smart scanning is utilized for detection of macro viruses that are written in text formats. It can ignore some characters employed to transform the appearance of the virus code, such as Space and TAB characters, and consequently improve the detection procedure quality, as a result.

### 3.2.2 Skeleton Detection

Skeleton detection is especially effective in order to detection of macro viruses. It does not utilize strings or checksums for detection purpose [14].

Eugene Kaspersky, Russian virus researcher and founder of the Kaspersky Anti-Virus, invented this technique and presented it for the first time. It reduces the searching zone inside a target file by removing each instruction that does not probably belong to the virus code before the scanning procedure starts. Firstly, the procedure parses statements of the macro virus one-by-one and removes any unimportant statements and all blanks gaps. Hence, the skeleton of the codes will be remained containing of only fundamental macro code, which the scanner exploit it to detect the virus [1].

### 3.2.3 Nearly Exact Identification

The purpose of nearly exact identification is more accurately detection of the viruses. One common method is to employ two strings as the signature of the virus, rather than only one. The virus is nearly exact identified, if both strings are existed in the file. Therefore, it makes disinfection process more reliable and risk-free, and ensures that the detected virus is not probable to be an unverified alternative of the primary version of the virus that maybe requires non- similar disinfection manner. Combination with bookmarks makes this technique more dependable.

Exploitation of a checksum range chosen from a virus code is also an alteration of nearly exact identification method that computes a checksum of the byte values in a specific area of the virus body. It brings about better accuracy, because a larger section of the virus body can be selected, with no need to overload the antivirus database.

In addition, it is not required to employ search strings, in order to implement nearly exact identification. In Kaspersky anti-virus algorithm, its creator, Eugene Kaspersky, does not make use of signature strings, instead employs two cryptographic checksums. These checksums are calculated at two specific locations, with given sizes inside the object.

### 3.2.4 Exact Identification

The exact identification technique utilizes non-variable bytes in the virus code as many as required to find a checksum of all bytes in the virus program, which contains constant value. The variable bytes of the virus body are ignored and a map of every constant byte is produced. This is the only method, which can promise an accurate detection of virus variants. It is often used as combination with the techniques of the first generation scanners. Exact identification method can discriminate exactly among various types of a virus, as well.

Even though it has many profits, but implementation of this technique make the scanners slow, slightly. In addition, really it is not easy to implement it for the outsized computer viruses.

### 3.2.5 Heuristics Analysis

The heuristics analysis is a useful method for detection of new unknown malwares [15]. It is especially helpful for detection of macro viruses too. It can be so worthwhile for binary viruses, as well, but it may extremely produce false positive output that is a major drawback of scanners [16].

Users cannot trust and will not purchase such anti-virus software that frequently produces tremendously false positives.

However, there are many situations, where a heuristic analyzer can be very valuable, and detect variants of a known virus family, as well. Heuristic analysis can be classified as two categories: static or dynamic [17]. Static heuristic is founded on the analysis of file structure and the code organization of the virus. While the static heuristic scanner is based on plain signs and code analysis to recognize the behavior of programs, the dynamic heuristic scanner performs CPU emulation of virus code, and tries to gather its information.

Some examples of heuristic flags are as following items, which express specific structural problems, may not be included in benign Portable Executables that are compiled using a 32-bit compiler [1]:
- Possible Gap between Sections
- Code Execution Starts in the Last Section
- Suspicious Section Characteristics
- Suspicious Code Section Name
- Virtual Size is Incorrect in Header of PE
- Multiple PE Headers
- Suspicious Imports from KERNEL32.DLL by Ordinal
- Suspicious Code Redirection

### 3.3 Virus-specific Detection

Sometimes the general virus detection algorithm may not be able to deal with a particular virus. In such conditions, a virus specific detection algorithm must be developed to carry out detection procedure. Actually, this kind of detection is not a regular method, but it denotes any special method that is specifically designed for a given particular virus. This approach is also called algorithmic scanning, but because it can be misleading [1], we use virus-specific detection term instead of algorithmic scanning.

This technique may bring about many problems such as portability of the scanner on different platforms and stability of the code. To overcome these problems, virus-scanning languages have been developed that in their plainest form, seeking and reading operations in scanned objects are allowed.

### 3.3.1 Filtering

This technique is used to optimize the performance of anti-virus engine regarding of scanning speed. It is especially useful in virus-specific detections because those are very time-consuming and high complexity in performance.

As a virus normally infects a particular or a set of known objects, signatures can be classified according to the infection type, such as .COM and .EXE files, boot sector, scripts, or macros, and so on. For example, executable viruses infect only programs such as .EXE and .COM, which are executable, macro viruses only attack to files or documents that can perform macro statements, and boot viruses place on the boot areas of disks. Through this exclusion action, when a specific file is searched for scan purpose, only the signatures relevant to its category are checked to keep scanning time down.

3.3.2 Static Decryptor Detection

As mentioned above, several types of viruses encrypt their body to prevent string scanning detection. In encrypted virus, the number of bytes, which can be used for string matching by scanners, is less. It makes trouble for string signature scanning engines. Therefore, anti-virus products have to use decryptor detection specific to a particular virus, which is not a very high quality method since it may produce many false negatives and false positives. In addition, because the virus body will not be decrypted during scanning, this technique cannot promise for a full disinfection.

However, the method can be a bit faster when it is employed together with an efficient filtering. It can also be employed to find other kinds of encrypted virus, such as oligomorphic or polymorphic viruses.

3.3.3 X-RAY Scanning

The X-RAY scanning method is also a virus-specific approach that is used to detect viruses of encrypted category, as well. X-ray scanning attacks the encryption of the virus rather than searching for the decryptor. It works based on a previously identified plaintext of the virus, and applies all encryption methods singly on special parts of files, such as top or tail of the file or supposed entry-point, to find the given plain text in decrypted virus body. X-raying exploits weaknesses of the virus encryption algorithm [18].This scanning method is able to find advanced polymorphic viruses, as well [1].

The following example from [18], explain a simple X-raying. One of the most common encryption functions usually used in viruses is XOR operation. Each byte of the encoded text is resulted by applying XOR function on a byte of the plain text with a fixed 8-bit value between 0 and *0xFF*, which is called the encryption key. For instance, consider a plain text *T = E8 00 00 5D* is encrypted with *XOR* operator and encryption key *k = 0x99*. After performing encryption, T will be converted as:
$$S = 71\ 99\ 99\ C4$$

We can decide whether *S* is an encrypted form of *T* or not, in two stages. Firstly, we can find the value of *k*, encryption key, a simple calculation using value of the first byte of *S*. In this example, if we assume that *71 = E8 XOR k*, then we can infer that *k = 0x99*. Secondly, we can check and verify whether the remaining bytes of the cipher text *S* can be decoded correctly using the assumed key *k* or not.

The weakness of this method is that it is very time-consuming when the start of the virus is not placed at a fixed location, so the encryption methods have to be applied on a large section of the file. The considerable advantage of this scanning is that it decrypts completely the virus body, and consequently makes disinfection possible, even the necessary information for removing is in encrypted form.

3.4 Code Emulation

This is one of the strongest detection techniques. It simulates the computer central processor, main memory, storage resources and some necessary functions of operating system by a virtual machine to run the malware virtually and investigate its behavior and performance. The malicious code does not execute on actual machine and it is controlled by the virtual machine precisely, therefore there is no risk for unintentionally propagation of malware.

The emulator imitates instructions of the machine by simulating CPU registers and flags, virtually. It resembles the execution of programs and detection procedure analyzes all instructions, individually.

For polymorphic viruses or other types of encrypted codes, after a given quantity of iterations or after a pre-defined stop situation, the scanner checks the contents of memory of the virtual machine. After sufficient iterations, polymorphic virus will decrypt its encrypted body and the real code will be revealed in the virtual memory. Scanner may use the following methods to choose when it breaks off the emulation loop: *Stopping with break points*, *Tracking of decryptor using profiles*, and *Tracking of active instructions*. When the emulation terminates, the virus will be checked by using string pattern matching or other scanning techniques [1].

Veldman in [19] called more generally this method as Generic Detection, in the case of any kind of encrypted malwares. He describes it as a way to decrypt an encrypted virus. Essentially, a generic detection consists of four parts: *processor emulator*, *memory emulator*, *system emulator*, and *decision mechanism* [2, 19].

Some more intelligent malwares alter their behavior or does not allow to be executed at all, if they perceive that there is an emulator. More about emulators and methods used to detect and attack emulators can be found in [20, 21, 22].

3.4.1 Dynamic Decryptor Detection

This is an attempt to detect the decryptor via emulation of the code. Actually, it is a method made of joining static decryptor detection and code emulation. It is helpful when the decryption loop is very long and time-consuming and code emulation merely is not suitable [1].

For example, it may identify the probable entry-point of the virus. Then, during the process, a specific algorithmic detection can examine the memory of virtual machines to find which areas have been modified. If it discovered any not reliable changes, extra scanning can verify the executed instructions and profile them, so the fundamental instructions set of decryptor loop can be recognized. Later this set can be exploited in order to detection of the virus. However, for the purpose of a perfect disinfection, because the complete decryption of the virus is required, the emulator has to simulate and execute the virus for a protracted time; accordingly, this procedure is not a practical approach [1].

For detecting more complicated polymorphic viruses, dynamic technique can be used, which employs code optimization procedure to make the decryptor routine smaller and transform it to a limited essential set of instructions by eliminating the dead code and non essential or garbage instructions, like NOPs and ineffective jumps that have no result. It helps to make emulation process faster and gives a signature for polymorphic decryptor.

## 4. Comparison

Table 2 summarized more common virus detection methods, which are explained above. Some more useful properties of detection methods are given in the table for a brief comparison. Symbol ✓ means the method can support the property or may affect on the property positively. Actually, symbol ✓ dedicates an advantage for the method, while symbol ✗ shows a weakness of the method. In scanning speed column, ✓ denotes that the method can improve the scanning speed and reduce the time complexity.

For example, from the table it can be seen that hashing techniques in first-generation scanners can improve the scanning speed and supports complete disinfection of the infected host, but it cannot used for detection of variants of a virus family, or unknown viruses or macro viruses. It has no effects on the false negative or false positive alarm, as well, in comparison to simple string signature scanning.

## 5 Conclusion and Future Recommendations

In this paper, though we try to review all most conventional antivirus techniques, but not all of them can be covered in a short survey.

Although the anti-virus software attempt to become updated and overcome the malwares threats, however we have to accept that virus authors are one step more ahead, because they decide how to attack first and anti-virus technologies have to only defense against their attacks. Therefore, computer virology area needs more researches and investigation to be able to guess the future coming threats.

There are many weaknesses in both viruses and anti-virus technologies, which must be studied and known well. Viruses usually look for the Achilles' heels in the defense system and attempt to attack them. Some major problems in detection methods are:

1  Most of detection methods are not powerful against evolutionary advanced or new viruses

2  Scanning process usually takes a considerable amount of time to search a system or networks for the patterns.

3  An anti-virus and its virus database need to be updated continually and extremely, otherwise it cannot be reliable.

So, interested researchers on the area of computer virology and anti-virus technologies are strictly recommended to work on these most important vulnerabilities.

Table 2: Comparison table of virus detection methods according to their features

IJCSI International Journal of Computer Science Issues, Vol. 7, Issue 6, November 2010
ISSN (Online): 1694-0814
www.IJCSI.org
120| | | | promise perfect disinfection | scanning speed improvement | virus family detection | new or unknown viruses detection | encrypted/polymorphic viruses | macro viruses | metamorphic viruses | false positive | false negative |
|---|---|---|---|---|---|---|---|---|---|---|---|
| first-generation scanners (string signature scanning) | | simple scanning | ✓ | ✗ | ✗ | ✗ | ✗ | ✗ | ✗ | Low | Low |
| | optimizing techniques | wildcards | ✓ | ✗ | ✓ | ✗ | ✗ | ✗ | ✗ | Low | Low |
| | | mismatch | ✓ | ✗ | ✓ | ✗ | ✗ | ✗ | ✗ | Low | Low |
| | | generic degree | ✓ | ✗ | ✓ | ✗ | ✗ | ✗ | ✗ | Low | Low |
| | | bookmarks | ✓ | ✗ | ✗ | ✗ | ✗ | ✗ | ✗ | Very Low | Low |
| | speed-up techniques | hashing | ✓ | ✓ | ✗ | ✗ | ✗ | ✗ | ✗ | Low | Low |
| | | top-and-tail scanning | ✓ | ✓ | ✗ | ✗ | ✗ | ✗ | ✗ | Low | High |
| | | entry-point/fixed-point | ✓ | ✓ | ✗ | ✗ | ✗ | ✗ | ✗ | Low | Low |
| second-generation scnners | | smart scanning | ✓ | ✓ | ✓ | ✗ | ✗ | ✓ | ✓ | Low | Low |
| | | skeleton detection | ✓ | ✗ | ✗ | ✗ | ✗ | ✗ | ✓ | Low | Low |
| | | nearly-exact identification | ✓ | ✗ | ✗ | ✗ | ✗ | ✗ | ✗ | Very Low | Very Low |
| | | exact-identification | ✓ | ✗✗ | ✓ | ✗ | ✗ | ✗ | ✗ | Zero | Zero |
| | | heuristic analysis | ✗ | ✗ | ✓ | ✓ | ✗ | ✓ | ✓ | Very High | Low |
| virus-specific detection | | general | ✓ | ✗✗ | ✓ | ✗ | ✓ | ✓ | ✓ | Low | Low |
| | optimizing techniques | filtering | ✓ | ✓ | ✗ | ✗ | ✗ | ✗ | ✗ | Low | Low |
| | | static decryptor detect | ✗ | ✗ | ✗ | ✗ | ✓ | ✗ | ✗ | Very High | Very High |
| | | X-RAY scanning | ✓ | ✗ | ✗ | ✗ | ✓ | ✗ | ✗ | Low | Low |
| code emulution | | Generic Detection | ✓ | ✗ | ✗ | ✗ | ✓ | ✗ | ✗ | Low | Low |
| | | dynamic decryptor detection | ✗ | ✓ | ✗ | ✗ | ✓ | ✗ | ✗ | Low | Low |

## References

[1] Szor, P., The Art of Computer Virus Research and Defense, Addison-Wesley Professional, 2005.

[2] Aycock, J., Computer Viruses and Malware, New York, NY, USA: Springer, 2006.

[3] Beaucamps, P., "Advanced Polymorphic Techniques", International Journal of Computer Science, Vol. 2, No. 3, 2007, pp. 194-205.

[4] Skulason, F., "Virus Encryption Techniques", Virus Bulletin, November 1990, pp. 13-16.

[5] Johansson, K., COMPUTER VIRUSES: The Technology and Evolution of an Artificial Life Form, 1994.

[6] Zhang, Q., "Polymorphic and metamorphic malware detection", Ph.D. Thesis thesis, ^Graduate Faculty, North Carolina State University, Raleigh, NC, USA, 2008.

[7] Bonfante, G., M. Kaczmarek, and J.Y. Marion, "Toward an Abstract Computer Virology", 2005.

[8] Ludwig, M., The Giant Black Book of Computer Viruses, Arizona: American Eagle Publications, 1995.

[9] Szor, P. and P. Ferrie, "Hunting for Metamorphic", in 11th Virus Bulletin International Conference, 2001, pp. 123-144.

[10] Szor, P., "The new 32-bit medusa", Virus Bulletin, December 2000, pp. 8-10.

[11] Jordan, M., "Dealing with Metamorphism", Virus Bulletin, October 2002, pp. 4-6.

[12] Vb, "IBM Viruses (Update)", Virus Bulletin, 1999, pp. 5-6.

[13] Erdogan, O. and P. Cao, "Hash-AV: Fast virus signature scanning by cache-resident filters", in IEEE Global Telecommunications Conference (GLOBECOM'05), 2005, Vol. 3, pp. 1767-1772.

[14] Catalin, B. and A. Vi Oiu, "Optimization of Antivirus Software", Informatica, Vol. 11, 2007, pp. 99-102.

[15] Bidgoli, H., Handbook of information security, Wiley, 2006.

[16] Arnold, W. and G. Tesauro, "Automatically generated Win32 heuristic virus detection", in 10th Virus Bulletin International Conference (VB2000), 2000, pp. 51-60.

[17] Nachenberg, C., "Understanding heuristics: Symantec's bloodhound technology", 1998.

[18] Perriot, F. and P. Ferrie, "Principles and practise of x-raying", in 14th Virus Bulletin International Conference (VB2004), 2004, pp. 51–56.

[19] Veldman, F., "Generic Decryptors Emulators of the future", in IVPC conference, 1998.
IJCSI
www.IJCSI.org

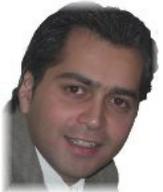

**Babak Bashari Rad** is currently a PhD candidate in Computer Science, at University Technology of Malaysia International Campus, Kuala Lumpur. He has completed his Master degree (2002) in Computer Engineering-Artificial Intelligence and Robotic as the outstanding student of Computer Science and Engineering (CSE) department, faculty of engineering at Shiraz University, Iran. He has been working as a faculty lecturer for nine past years in Azad University branches at Iran. His interests and research areas are computer virology, malwares, information security, code analysis, and machine learning methodologies. He is studying on metamorphic virus analysis and detection methodologies for his PhD thesis.

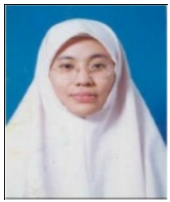

**Maslin Masrom** received the Bachelor of Science in Computer Science (1989), Master of Science in Operations Research (1992), and PhD in Information Technology/Information System Management (2003). She is an Associate Professor, Razak School of Engineering and Advanced Technology Malaysia, Universiti Technology Malaysia International Campus, Kuala Lumpur. Her current research interests include information security, ethics in computing, e-learning, human capital and knowledge management, and structural equation modeling. She has published articles in both local and international journals such as Information and Management Journal, Oxford Journal, Journal of US-China Public Administration, MASAUM Journal of Computing, ACM SIGCAS Computers & Society and International Journal of Cyber Society and Education.

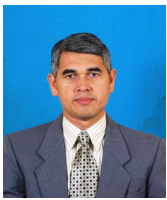

**Suhaimi Ibrahim** received the Bachelor in Computer Science (1986), Master in Computer Science (1990), and PhD in Computer Science (2006). He is an Associate Professor attached to Advanced Informatics School (AIS), Universiti Teknologi Malaysia International Campus, Kuala Lumpur. He is an ISTQB certified tester and currently being appointed a board member of the Malaysian Software Testing Board (MSTB). He has published articles in both local and international journals such as the International Journal of Web Services Practices, Journal of Computer Science, International Journal of Computational Science, Journal of Systems and Software, and Journal of Information and Software Technology. His research interests include software testing, requirements engineering, Web services, software process improvement and software quality.